\documentclass{article}

\usepackage{graphicx}
\usepackage{amssymb}

\newcommand{\code}[1]{\texttt{#1}}
\begin{document}

\title{How to turn a scripting language into a domain specific language for computer algebra}

\author{
Raphael Jolly\\
{\small Databeans}\\
{\small Paris, France}\\
\texttt{raphael.jolly@free.fr}
\and 
Heinz Kredel\\
{\small IT-Center, University of Mannheim}\\
{\small Mannheim, Germany}\\
\texttt{kredel@rz.uni-mannheim.de}
}

\maketitle

\begin{abstract}
  We have developed two computer algebra systems, meditor \cite{Jolly:2007}
  and JAS \cite{Kredel:2006}. These CAS systems are available as Java libraries.
  For the use-case of interactively entering and manipulating mathematical
  expressions, there is a need of a scripting front-end for our libraries.
  Most other CAS invent and implement their own scripting interface
  for this purpose. We, however, do not want to reinvent the wheel and
  propose to use a contemporary scripting language with access to Java
  code. In this paper we discuss the requirements for a scripting language
  in computer algebra and check whether the languages Python, Ruby,
  Groovy and Scala meet these requirements. We conclude that, with
  minor problems, any of these languages is suitable for our purpose.
\end{abstract}

%
%

\section{Introduction} 

In this paper we summarize the numerous discussions that resulted from
our encounter as developers of two computer algebra systems (CAS)
written in Java \cite{Jolly:2007,Kredel:2006} that have grown
independently during the past 7 years (roughly). The focus of this
paper is the scripting requirements for a computer algebra system. We
compare various design choices and solutions and conclude that modern
scripting languages are suitable for our tasks. We first give some
background information on computer algebra systems design and discuss
the scripting requirements in section 2. In section 3 we investigate
whether modern scripting languages are suitable for our needs. In the
next section we give an overview of the state of our software and then
conclude with our findings. Although we focus on computer algebra the
topic of this paper is relevant to all kinds of interactive scientific software
which want to allow users to input mathematical expressions.

After the construction of the first mature programming languages, like
FORTRAN or later C, thousands of program libraries have been developed.
Moreover, the popularity of a programming language is often based on
the portfolio of available algorithms in the programming libraries.
The quality and extent of the library is considered so important, for
example for current scripting languages like Perl, PHP or Ruby, that these
come with built-in support for online access and easy installation of new
program modules by means of CPAN, Pear or Gem facilities. Also Java's
success is influenced by its standardized and comprehensive libraries
on which every programmer can depend when writing and deploying
their own programs.

In computer algebra the situation is quite different. Java was the
first programming language to encompass the requirements for
the implementation of algebraic algorithms in their full diversity :
interfaces for the specification of algebraic structures and dynamic
data-structures with run-time support for it (for example garbage
collection). Of course, Lisp allows dynamic data-structures, but
because of its sloppy typing system, it seems unsuitable for a CAS
library. Second, the user-interface requirements for writing down
arbitrary expressions in the desired algebraic structures hinder the
success of algebraic libraries. Most computer algebra software has
some form of an algebraic expression language with an interpreter
and some GUI support to edit and transform the expressions. Actual
programming languages (like Java) cannot fulfill these requirements,
the construction of elements of algebraic structures, although possible,
is tedious and far away from the paper-and-pencil form of expressions.

As we have shown with JSCL and JAS, the first aspect of algebraic
libraries has been solved for Java, others have established libraries
in C++, like GiNaC \cite{Kreckel:2002}. For the second point the
situation is only slowly changing. Improvements are seen around
scripting languages, notably the Python language was the first to
demonstrate facilities to allow the nearly paper-and-pencil form of
expressions to be entered \cite{Certik:2008,Stein:2005}. In this paper,
we investigate current scripting languages for their suitability for the
user interface requirements of computer algebra systems. We do not
discuss issues around graphical user interfaces and editor
components for computer algebra.

\subsection{Algebraic libraries} 

Historically, computers were first built to automate numerical tasks. Since
then, numerous attempts were made to supplement these with some symbolic
capabilities. The language of choice for that purpose, as inferred from the
resulting products \cite{Grabmaier:2003}, is either C (Mathematica, Maple)
or LISP (Macsyma/Maxima, Axiom, Reduce). There are attempts in C++
(MuPAD), but they are not as advanced. Despite the advantages of object
oriented programming, programmers working in natural science (for
example physics) object against using OOP. One reason is the perceived
better performance of procedural languages, say FORTRAN. Another
reason is that there exist large well tested libraries, developed over
decades, which would have to be recoded in an OOP language. So
we see OOP code mainly in new developments or as `glue' code to
tie together legacy codes. In computer algebra there is not much
published on object oriented algorithm implementation. There is a
first paper on CAS with SmallTalk \cite{ AbdaliCherrySoiffer:1986} and
the ongoing work of the Axiom developers can, to some extent, be viewed
as object oriented \cite{JenksSutor:1992}. Newer approaches start with
\cite{Zippel:2001} in Common Lisp, then using C++
\cite{Kreckel:2002,Parisse:2008}. Approaches using Java are
\cite{Niculescu:2003,Niculescu:2004,Whelan:2003} and
\cite{Platzer:2005,Jolly:2007,Kredel:2006,Kredel:2007,Kredel:2008}.

\subsection{Scripting languages} 

In parallel to the aforementioned symbolic packages, a means for user
interaction was also researched, which resulted in the widely adopted
form of the command line interface (CLI). Some projects however left
the concern aside, and centered on library development alone. In that
respect, GiNaC for instance can not be used interactively. It can only
be called from other programs, which is a bit frustrating. Command
line interfaces on the other hand usually come with a corollary
scripting capability (interpreter).

The latter is often used in turn to develop higher-level extension
libraries. The drawback to this approach is that the system ends up
being developed in two (or more) different languages, one being
usually a general purpose, compiled language, for the kernel and
core libraries, and the other(s) (an) interpreted, domain specific,
language(s), for the extensions. As a result, there are as many
different, obviously incompatible extension languages for algebra,
as there are products.

Hence today, computer algebra systems universally resort in some
extent to domain specific languages (DSL). This is unfortunate,
because it implies the additional burden of setting up a whole
grammar, parsing mechanisms, etc. Even in projects like Xcas/Giac
\cite{Parisse:2008} that seek to eliminate a specific language for
extensions, leveraging the features of the host language instead
(operator overloading etc.) there is the need for parsing user input.

An attempt to counter the multiplication of extension languages is
made by the Sage project \cite{Stein:2005}, which aims to unite
several libraries in the fold of one scripting interface. But there is
still the dichotomy between the script, which is Python, and the
various pieces of (compiled) code. As a matter of fact, there are
plans to rewrite the engine in Python \cite{Certik:2008}.

The Java platform could bring a solution based on its scripting
framework (JSR-223) : one could take one of its existing, widely used,
general purpose scripting languages to handle user inputs. As it
happens, there are increasingly many available : Beanshell, Rhino,
Jython, JRuby, Groovy, Jaskell (for Python, Ruby, Groovy and Scala see
\cite{vanRossum:1991,Matsumo:1995,Groovy:2003,Odersky:2003}).
Beanshell is used by meditor. JAS on its hand uses Jython.

\section{Desired language features for computer algebra} 
\label{sec:req}

The interactive use of a CAS mainly consists of entering algebraic
expressions and calling methods to compute a desired result or to
transform the expressions to some other forms. For example in
algebraic geometry one could want to enter

\begin{verbatim}
  cas> f = x**3 * y + x**2 * z - 5/9
  cas> g = y**4 - z**6 + 7 * w
  cas> F = [ f, g ]
  cas> G = groebner( F, graded )
  cas> I = ideal( G ); J = ...
  cas> K = I.intersect( J )
\end{verbatim}

Here, \code{f} and \code{g} are variables, which get assigned
polynomial literals. A list \code{F} is constructed from the
polynomials and a Gr\"obner base \code{G} is computed.
From these parts, ideals \code{I}, \code{J} are constructed
and an ideal intersection \code{K} is computed.


In this example the first two lines are the greatest problem for the
scripting language. The remaining lines can easily be achieved in almost
any scripting lanuage, perhaps with some different `syntactic sugar'.
The first problem with polynomial literals (or similar expressions)
comes from the use of variables \code{x}, \code{y} and \code{z} which
have no assigned value and no definition in the history of the script
execution. The second problem is the use of language operators like
\code{*}, \code{+} or \code{**} on a mixture of number literals and
variable literals. A third problem arises from the use of operators
like \code{/} in \code{5/9} on number literals which is ambiguous, as
it is not clear if we mean integer division, floating point division or
even creation of rational numbers. Finally the software must understand
that \code{f} is a polynomial in the ring $\mathbb{Q}[x,y,z]$ and \code{g}
is a polynomial in the ring $\mathbb{Z}[w,x,z]$, and eventually that \code{F}
must be a subset of the ring $\mathbb{Q}[w,x,y,z]$. Subsequent
computations would then take place in the last ring.

In the rest of this section we study the requirements on a scripting
language which arise from such computations : the definition of
symbols (variable literals), operator usage and coercions.


\subsection{Definition of symbols} 
\label{sec:sym}

An important need relates to syntax. Writing polynomials natively
(without quotes) in usual notation, like \verb/1+x+2*x^2/, is a must
for computer algebra. Currently, not many languages can do that
acceptably. A symbol like \code{x} here will be an object in the sense
of OOP. In \cite{Certik:2008} for instance it is declared as follows:

\begin{verbatim}
  >>> from sympy import *
  >>> x,y = symbols('x', 'y')
\end{verbatim}

This initializes the language variables `x' and `y' to objects of class
\code{Symbol}. Additionally all single letter variables and some greek
letter names are predefined during SymPy initialization. Note, however,
there is no way to avoid self-confusion if one later assigns it to a variable
with different name.

\begin{verbatim}
  >>> z = x
\end{verbatim}

Now `z' points to a symbol with name `x'. In Sage \cite{Stein:2005} a
polynomial ring declaration and polynomial expression is made as:

\begin{verbatim}
  sage: R.<x> = PolynomialRing(QQ,1); R
  sage: f = (x+1)^20
\end{verbatim}

We see that, contrary to SymPy, Sage needs the ring to be known
before variables are instantiated. We envision the same kind of
declaration as in these two projects.

Hence, in conjunction with the next point `operator overloading' we
will be able to build expressions from number literals, symbols, function
names and operators. These expressions will in turn be able to be used
in further arithmetic constructs, converted to other types, and so on.

\subsection{Operator overloading} 

We want to use operator overloading to emulate the way arithmetic
operators act on numeric types, this time for symbolic types. In a scripting
language, it is required in two places. First, in expressions containing
symbols, the evaluation must be suppressed when an operator encounters
a symbol as operand. Second, when expressions contain (sub-)expressions
involving objects form the Java libraries, the operation must use the
respective class methods.

For example in \code{2*3+x}, the `\code{+}' operation may not try to
operate on symbol `\code{x}', but \code{2*3} should be evaluated and
simplified to \code{6}. Or, if \code{f} and \code{g} are of type
\code{Multivariate\-Polynomial}, an expression like \code{f-g} should
use the \code{subtract()} method of the respective class and simplify
the expression to \code{f.sub\-tract(g)}.

How exactly operator overloading must be implemented is debated.
There are different levels of language support, depending on whether
it can be customized, what operators can be used, and so on. There is
also the problem of precedence.

Java has limited support for operator overloading: \verb/+/ is used
both for numeric types and \code{String}s. Some proposed changes to
the JDK consist in extending it to \code{BigInteger} and
\code{BigDecimal} (as in Groovy), or to a new interface (e.g.
\code{Arithmetic}) that would be implemented by \code{java.math.*} and
any other user defined classes like \code{Complex}, \code{Matrix}, and
so on.

An important problem relates to the use of \verb/^/ for \code{pow}.
Because of the legacy of C, where it is used for exclusive or, nearly all
modern language have the atavism of not being able to use it for this
purpose. Even when it can be overloaded, it doesn't have correct
precedence \cite{Parisse:2008}.

There is a similar problem regarding rationals, like \code{1/2} which is
evaluated to \code{0} in many existing languages. Or even worse, the
standard behavior of \code{/} is redefined to mean floating point
division and to return \code{0.5}.

\subsection{Conversions / Coercions} 
\label{sec:cnv}

Suppose we have to find an implementation for an operator on
incompatible algebraic types. In Perl, the operator would force the
operands to be converted as required. In strongly typed object oriented
languages we face a problem. In the Java library code we simply require
all types to be defined at compile time, but in the interpretation of a script
this seems overly restrictive. During script execution there should be an
attempt to find an algebraic structure where the operands can be
embedded into and the operator has a meaningful implementation.

In Axiom \cite{JenksSutor:1992}, if no information is given, then first
the most general type of the two operand expressions is searched.
Then both expressions are transformed to this `bigger' type and then
the method is called. So for the polynomial ring $K[x]$ as a set:
it contains $x$ and also all elements of $K$. For $1+x$ the
system would deduce that $1$ is in $K$ and that the bigger
set which also contains $x$ is then $K[x]$.

This is difficult to deduce from object oriented expressions.
For the expression \code{1.add(x)} one must find out that the result is
in $K[x]$, represented by some object, say \code{univPolyX} and rewrite
the expression to \code{univPolyX.valueOf(1).add( uniPolyX.valueOf(x))}.
In a similar way one could handle $x+y$ as \code{x.add(y)} and rewrite
it to \code{polyXY.valueOf(x).add( polyXY.valueOf(y))}. The latter
supposes among other things that the variable names are at hand.

A different approach would defer the evaluation of the expression
\code{1.add(x)} until more type information is available or provided
by the user. Also Axiom has a notation to explicitly request a type for
an expression and Sage is going the same way in requesting the
specification of a polynomial ring before objects from the ring are
entered in expressions, as we have noted (section \ref{sec:sym}).

We must pay special attention to the case where one or the other operand
is of a numeric type. The statement \code{a+b} will be internally translated
into either \code{a.add(b)}, which is a unary instance call, or \code{add(a,b)},
which is a binary static call, depending on how operator overloading is implemented.
In the first option, there is a problem when \code{a}, which is called the `target', is
numeric, that is, a primitive type and not a true object. Then we must either have
it converted, if possible implicitly, to a reference type, which additionally must be
able to be overloaded, or to swap it with \code{b}, which is called the `argument',
putatively a reference type here. The latter mechanism is called `double dispatch'.

\section{Assessment of script\-ing lang\-uages} 
\label{sec:ass}

\begin{figure}
\begin{center}
\includegraphics[clip,width=0.8\textwidth]{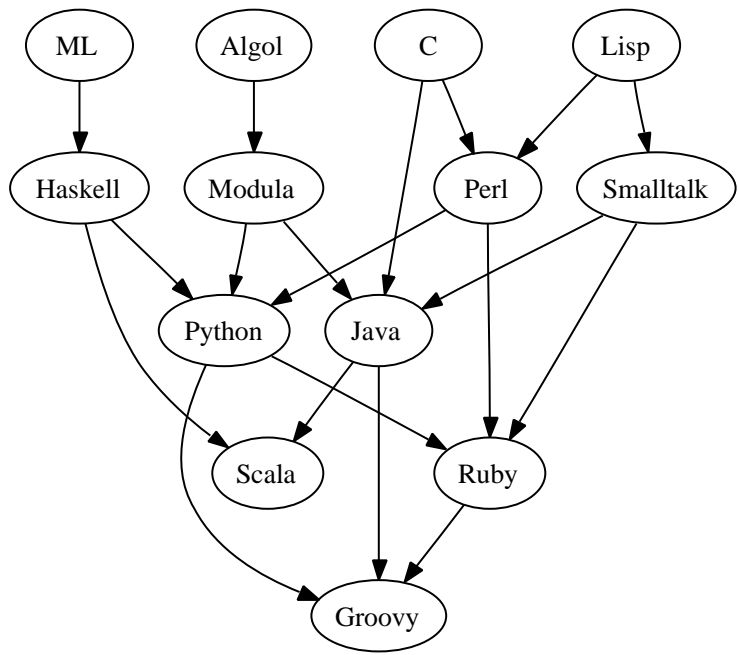}
\caption{Genealogy of scripting languages}
\label{fig:genealogy}
\end{center}
\end{figure}

With the list of requirements from section \ref{sec:req}, we
now examine whether modern scripting languages (see figure
\ref{fig:genealogy} for a genealogy) can be used to implement a
scripting front-end to a CAS. We focus on languages which have an
implementation which can access Java library code, such as Python
with Jython \cite{vanRossum:1991,Jython:1997}, Ruby with JRuby
\cite{Matsumo:1995,JRuby:2003} and the systems Groovy
\cite{Groovy:2003}, Scala \cite{Odersky:2003} which are
directly implemented with Java code access.

The scripting language moreover must support some kind of strong
typing, as otherwise objects or methods from the back-end Java
libraries of the wrong type could be used. Strong typing means that
constructed objects have types which define the allowed methods of the
object. Using undefined methods or parameters with wrong types is
considered a type error and an exception is thrown.

There are more issues to consider when selecting a scripting
language, such as performance, availability, graphical user
interface capabilities or size of their community. We do not
compare the languages with respect to these issues in this paper.

In order to compare the different languages, we define a toy
application that tries to capture the key issues of a real CAS
implementation. We try to setup classes, so that we can print
the expression `\code{1+x}' from the previous section, with only
declaring `\code{x}' to be a symbol, like `\code{x = Sym('x')}'.

The main classes in the following examples are a \code{Sym} class
representing symbols and an \code{Expr} class representing arbitrary
expressions. In both classes we let the \code{+} operator return an
\code{Expr}. For \code{Sym}, the `to string' method returns the string
name of the symbol and for \code{Expr} the `to string' method returns
the concatenation of the left and right operands with the operator in
the middle. The other operators \code{*}, \code{-}, \code{/} etc. can
then be implemented similarly. For function representation, such as
\code{sin(x)} we will use a \code{Func} class, which holds the name
of the function and expressions of all arguments.

\subsection{Python} 

The code of \code{Expr} and \code{Sym} for Python follows.
The method \verb/__add__/ is called when the interpreter
encounters a \code{+} operation on symbols or expressions.
\verb/__radd__/ interchanges (reverses) the operands of \code{+}.
The \verb/__str__/ method is like Java's \code{toString()} and
\verb/__init__/ is Python's constructor method.

\begin{verbatim}
class Expr:
   def __init__(self,left,op,right):
       self.left=left
       self.op=op
       self.right=right
   def __add__(self,other):
       return Expr(self,'+',other)
   def __radd__(self,other):
       return Expr(other,'+',self)
   def __str__(self):
       return str(self.left)\
              +str(self.op)\
              +str(self.right)

class Sym:
   def __init__(self,name):
       self.name=name
   def __add__(self,other):
       return Expr(self,'+',other)
   def __radd__(self,other):
       return Expr(other,'+',self)
   def __str__(self):
       return str(self.name)

print(1+2) # --> 3
x=Sym('x'); y=Sym('y')
print(1+x) # --> 1+x
print(x+y) # --> x+y
\end{verbatim}

In this sample, we instantiate two arithmetic objects, we add them,
and we print the result. This is the way real world objects like
polynomial or matrix would behave in our envisioned setting.
We see that the operator overloading implementation is perfectly
suited to our purpose (\verb/__add__/ method). Regarding what
happens if the target is numerical, Python uses double-dispatch
(\verb/__radd__/ method).

Regarding the power issue, Python uses \code{**}. The Sage crew have
addressed the problem by preparsing the input to convert from \verb/^/
to \code{**}.

For fractions, there has been talk in the Python community
about changing Python so \code{2/3} returns the floating point number
\code{0.6666...}, and making \code{2//3} return \code{0}, which is of
no use as far as computer algebra is concerned.

\subsection{Ruby} 

The Ruby language and its interpreter is capable of redefining built-in
classes and implementing methods for coercion between classes. In
Ruby, numbers are objects, so \code{1+2} is an abbreviation of
\code{1.+(2)}. \code{1.class} returns \code{Fixnum}, and \code{+}
is a valid method name, which can be overridden (not overloaded).

\begin{verbatim}
class Expr
  def initialize(left,op,right)
      @left=left; @op=op; @right=right
  end
  def +(other)
      Expr.new(self,'+',other)
  end
  def coerce(other)
    [Num.new(other),self]
  end
  def to_s
    "#{@left}#{@op}#{@right}"
  end
  attr_reader :left, :op, :right
end

class Sym
  def initialize(name)
      @name=name
  end
  def +(other)
      Expr.new(self,'+',other)
  end
  def coerce(other)
    [Num.new(other),self]
  end
  def to_s
    "#{@name}"
  end
  attr_reader :name
end

class Num
  def initialize(value)
      @value=value
  end
  def +(other)
      Expr.new(self.value,'+',other)
  end
  attr_reader :value
end

puts 1+2   # -> 3
x = Sym.new('x'); y = Sym.new('y')
puts 1+x   # -> 1+x
puts x+y   # -> x+y
\end{verbatim}

When Ruby encounters an expression like \code{1+x} and cannot find
a suitable method to add a \code{Sym} to a \code{Fixnum} it calls a
method \code{coerce(1)} on the \code{Sym} object. If we implement
\code{coerce} as wrapping \code{Fixnum} in class \code{Num}, which
is aware of our expression tree, we get the desired effect. This is sort of
a mix between Python's \verb\__radd__\ and Scala's implicit conversion
(see below). The resulting expression tree can then be printed via its
\verb/to_s()/ methods or be transformed in any desired way. The method
\code{initialize()} is the constructor method. \verb/@name/ denotes
instance variables and \verb/#{@left}/ denotes substitution of the
variable \verb/@left/ in a string.

Instead of using the \code{coerce} method we could also extend the
built-in class \code{Fixnum} with the following implementation.

\begin{verbatim}
class Fixnum
  alias plus_old +
  def +(other)
      if other.class == Sym or other.class == Expr
         Expr.new(self,'+',other)
      else
         self.plus_old(other)
      end
  end
end
\end{verbatim}

In this case we redefine the method \code{+()} of the built-in class
\code{Fixnum} to respect our new symbol and expression classes
\code{Sym} and \code{Expr}. If \code{+()} encounters one of them, it
constructs a new \code{Expr} node. The old meaning of \code{+} is
remembered as an alias, \code{plus\_old}, and is called on
\code{Fixnum}s. The other classes where \code{+} must be overridden
are \code{Bignum}, \code{Rational}, \code{Float}, \code{BigDecimal}
and \code{Complex}.

Lastly, regarding the fraction problem, Ruby defines `\code{/}' to mean
construction of a rational number when the module `\code{rat\-ional}'
is loaded.

\subsection{Groovy} 

In Groovy, the method \code{plus()} is called when the interpreter
encounters a \code{+} operation.

\begin{verbatim}
class Expr {
  Object left
  String op
  Object right
  Expr plus(Object other) {
    return new Expr(left:this,op:"+",right:other)
  }
  String toString() {
    return left.toString()+op+right.toString()
  }
}

class Sym {
  String name
  Expr plus(Object other) {
    return new Expr(left:this,op:"+",right:other)
  }
  String toString() {
    return name
  }
}

class Arithmetic {
  static Expr plus(Integer a, Object b) {
    return new Expr(left:a,op:"+",right:b)
  }
}

use(Arithmetic) {
  println(1+2) // -> 3
  x = new Sym(name:"x"); y = new Sym(name:"y")
  println(1+x) // -> 1+x
  println(x+y) // -> x+y
}
\end{verbatim}

In this language, numbers are objects, so \code{1.class} returns
\code{java.lang.Integer} and \code{1+2} is a shortcut of
\code{1.plus(2)}. If we want the second operand to be non-numerical,
we have to resort to the \code{use(Class\-Name)} construct, as is done
in \cite{JScience:2006}. \code{Arithmetic} is a so called `category class',
a class with only static method definitions. Within the \code{use} block
the interpreter uses the method \code{plus} from \code{Arithmetic}
when a \code{+} operation on numbers and \code{Object}s is
encountered.

Another option in Groovy is to use the `ExpandoMetaClass' construct:

\begin{verbatim}
Integer.metaClass.plus = {
  Integer other -> delegate-(-other)
}

Integer.metaClass.plus = {
  Object other -> new Expr(left:delegate,op:"+",right:other)
}
\end{verbatim}

This avoids the `use' block. Here we have used a subtract
operator trick, but in a real implementation it would require to
re-implement original operators in low-level java.

Regarding the fraction issue, Groovy computes \code{1/2} as a float
\code{0.5}. It also computes a BigInteger to some power as BigDecimal
\code{2G**200 $\longrightarrow$ 1.6069380442589903E60}.
Obviously these choices were made with no symbolic or algebraic
computation concern (though the suffix \code{G} for BigInteger
literals could be useful). It is however possible to overload the div
operator as seen above, so that \code{1/2} does what we want.

\subsection{Scala} 

In Scala, \code{+} is a valid method name, so it is used to define
the addition of objects. Here is how the sample is coded in Scala:

\begin{verbatim}
class Expr(left:Any,op:String,right:Any) {
  def +(other:Any)=new Expr(this,"+",other)
  override def toString=left+op+right
}

class Sym(name:String) {
  def +(other:Any)=new Expr(this,"+",other)
  override def toString=name
}

class Num(value:Int) {
  def +(other:Any)=new Expr(value,"+",other)
}

implicit def int2num(n: Int): Num = new Num(n)

println(1+2) // -> 3
val (x,y) = (new Sym("x"),new Sym("y"))
println(1+x) // -> 1+x
println(x+y) // -> x+y
\end{verbatim}

To be able to mix numeric operands, we have to use the `view' mechanism
(\code{implicit} keyword) which is a generalization of auto-(un)boxing in
Java. Based on this mechanism the expression \code{1+x} will be rewritten
as \code{int2num(1)+x}.

Regarding the power issue, \verb\**\ has a wrong precedence (same as
\verb\*\, based on the first character), and the only possible single ASCII
characters are \verb/?/ and \verb/\/, which shows that computer algebra
was not taken into account in the design of the language (until now).
Unicode can be used though, which would allow special chars as
`up arrow' (and the Greek alphabet, but this is a different matter).

Regarding the fractions, they could be handled with the 1\verb\%%\2
syntax, with \verb\%%\ an operator on a symbolic type, which would
return a \code{Frac(1,2)} object for instance.

\section{State of current software and future work} 

In the designs of our computer algebra systems, meditor and JAS, we
have been using Beanshell and Jython as scripting front-ends to the
algebraic Java libraries.

With Jython, we provided the scripting classes \code{Ring} and
\code{Ideal}. \code{Ring} is an interface to the Java class
\code{Gen\-Polynomial\-Ring}. Its string constructor is passed to a
Java string tokenizer, which returns a polynomial ring object. Given
the ring object, it is possible to construct a polynomial ideal
\code{Ideal} in this ring from a string representation, also via a
Java string tokenizer. So the example from section \ref{sec:req}
reads in JAS as:

\begin{verbatim}
  jas> r = Ring("Q(w,x,y,z) G")
  jas> I = r.ideal("( (x**3 y + x**2 z - 5/9),\
                   (y**4 - z**6 + 7 w) )" )
  jas> I = I.GB(); J ...
  jas> K = I.intersect( J )
\end{verbatim}

Regarding meditor, the advantage of Beanshell is that its syntax is
almost identical to Java. This is also its drawback since it lacks the
features that we have discussed. The way that meditor deals with
the example is as follows:

\begin{verbatim}
  bsh % v = Variable.valueOf(new String[]
        {"w", "x", "y", "z"});
  bsh % a = new Generic[] {
        Expression.valueOf("x^3*y + x^2*z - 5/9"),
        Expression.valueOf("y^4 - z^6 + 7*w")};
  bsh % a = Basis.compute(a, v,
        Monomial.totalDegreeLexicographic,
        1).elements(); // 1 is for Q
  bsh % // ideal intersection not implemented
\end{verbatim}

In both programs, polynomials are represented as strings instead of
expressions in the host language. Parsers had to be implemented and
maintained and have fewer possibilities to spot errors in the denotation
of polynomials. This is very rudimentary compared to the possibilities
of the scripting languages investigated in this paper.

Note, with our custom parsers we could allow the multiplication
operator `\code{*}' to be optional, which seems to be impossible
with scripting languages.


For future development we have to decide whether we follow the
`nested expression' approach (based on class \code{Expr}) used in our
toy sample code or if we implement the nested expression in our back-end
Java libraries or if we directly use suitable existing polynomial classes
from the library. That is, instead of defining new classes \code{Sym}
and \code{Expr} in our scripting language we could reuse existing Java
classes like \code{Expression}, or \code{PolynomialRing} or implement
classes similar to \code{Sym} and \code{Expr} directly in Java.

The \code{Expression} class of meditor is a kind of polynomial capable
of extending variables dynamically, so it could directly be substituted for
\code{Expr}. For JAS, there are only fixed variable number polynomials
\code{GenPolynomial} which need a polynomial ring factory before any
of them can be created. In this case we can implement
\code{collectSymbols(expr)} and \code{mostGeneralNumberType(expr)}
methods which will give a list of all symbols occurring in an expression
and the maximal type for the coefficients. With this we can implement a
polynomial or ring constructor, as in:

\begin{verbatim}
  cas> r = PolynomialRing( [ expr, ... ] )
  cas> p = Polynomial( expr )
\end{verbatim}

Alternatively, we could define the algebraic structure by hand and use
a factory method, like \code{valueOf()}, to coerce expressions to the
desired type:

\begin{verbatim}
  cas> r = PolynomialRing(Q,'w,x,y,z')
  cas> F = [ r.valueOf(f), r.valueOf(g) ]
\end{verbatim}

We will investigate these issues in future implementation studies
with any of the scripting languages. As a side condition we will try
to maintain some kind of compatibility with SymPy and Sage.

\section{Conclusion} 

For the interactive use of our computer algebra systems there is a
need of a scripting front-end for the Java libraries. We propose to
use a contemporary scripting language with access to Java code. The
requirements for a scripting language in computer algebra is to be
able to define symbols, override arithmetic operators and the ability
to extend built-in classes. The languages Python, Ruby, Groovy and
Scala have been selected because they all can call Java library code
and are strongly typed.

Working with simplified toy code we showed how to implement the
required features. All languages can define symbols and can override
arithmetic operators.

With Python one can define arithmetic operators with reversed
arguments to meet our requirements. So we can achieve our goals with
the least effort. We only have to implement \verb\__r*__\ methods for
all classes we want to interact with.

Ruby can use explicit coercion methods for all kinds of unknown
methods for built-in number classes. So to achieve our goals we have
to implement an expression tree aware wrapper class for numbers and
to implement \code{coerce()} methods for our new expression tree classes.
Alternatively Ruby can extend built-in classes to add new methods or
override existing methods.


Groovy can specify a context where an arithmetic operator is used
with a desired implementation. Since a context requires to be enclosed
in braces, users would have to type a closing brace before the code
could be interpreted. Alternatively, build-in classes can be extended,
but this requires a re-implementation of original operators in low-level
java. To reach our goals we have to implement overloaded methods for
all classes we want to interact with.

In Scala we can define global methods to achieve desired object
coercions (transformations) to our data types. Scala's view mechanism
detects a missing connection and finds the global method according
to the required method signature. So to reach our goals we have to
implement global coercion methods for all classes we want to interact
with. Most of the work is then done via implicit conversion.

The extension of built-in classes may add a performance penalty
compared to context specific overriding, which we have not studied.


To conclude, with minor problems any of these languages is suitable
for our purpose.

\subsection*{Acknowledgments} 

We thank our colleagues for various discussions and for encouraging
our work.


\bibliographystyle{abbrv}
\bibliography{scripting}

\begin{thebibliography}{10}

\bibitem{AbdaliCherrySoiffer:1986}
S.~K. Abdali, G.~W. Cherry, and N.~Soiffer.
\newblock An object-oriented approach to algebra system design.
\newblock In B.~W. Char, editor, {\em Proc. SYMSAC 1986}, pages 24--30. ACM
  Press, 1986.

\bibitem{Certik:2008}
O.~Certik.
\newblock {SymPy} {Python} library for symbolic mathematics.
\newblock Technical report, http://code.google.com/p/sympy/ accessed, April
  2008, since 2006.

\bibitem{JScience:2006}
J.-M. Dautelle.
\newblock {JScience}: {Java} tools and libraries for the advancement of
  science.
\newblock Technical report, http://www.jscience.org/, accessed 2007, May,
  2005-2007.

\bibitem{Kreckel:2002}
A.~Frink, C.~Bauer, and R.~Kreckel.
\newblock Introduction to the {GiNaC} framework for symbolic computation within
  the c++ programming language.
\newblock {\em J. Symb. Comput.}, 2002.

\bibitem{Grabmaier:2003}
J.~Grabmaier, E.~Kaltofen, and V.~Weispfenning, editors.
\newblock {\em Computer Algebra Handbook}.
\newblock Springer, 2003.

\bibitem{Groovy:2003}
{Groovy Developers}.
\newblock {Groovy} an agile dynamic language for the {Java} platform.
\newblock Technical report, http://groovy.codehaus.org/, accessed 2008, April,
  2003-2008.

\bibitem{vanRossum:1991}
{Guido van Rossum}.
\newblock {Python} a dynamic object-oriented programming language.
\newblock Technical report, http://www.python.org/, accessed 2008, April,
  1991-2008.

\bibitem{JenksSutor:1992}
R.~Jenks and R.~Sutor, editors.
\newblock {\em {axiom} The Scientific Computation System}.
\newblock Springer, 1992.

\bibitem{Jolly:2007}
R.~Jolly.
\newblock jscl-meditor - java symbolic computing library and mathematical
  editor.
\newblock Technical report, http://jscl-meditor.sourceforge.net/, accessed
  2007, September, since 2003.

\bibitem{JRuby:2003}
{JRuby Developers}.
\newblock {JRuby} a {Java} powered {Ruby} implementation.
\newblock Technical report, http://jruby.codehaus.org/, accessed 2008, April,
  2003-2008.

\bibitem{Jython:1997}
{Jython Developers}.
\newblock {Jython} implementation of the high-level, dynamic, object-oriented
  language {Python} written in 100\% pure {Java}.
\newblock Technical report, http://www.jython.org/, accessed 2008, April,
  1997-2008.

\bibitem{Kredel:2006}
H.~Kredel.
\newblock On the {Design} of a {Java} {Computer} {Algebra} {System}.
\newblock In {\em Proc. PPPJ 2006}, pages 143--152. University of Mannheim,
  2006.

\bibitem{Kredel:2007}
H.~Kredel.
\newblock Evaluation of a {Java} {Computer} {Algebra} {System}.
\newblock In {\em Proceedings ASCM 2007}, pages 59--62. National University of
  Singapore, 2007.

\bibitem{Kredel:2008}
H.~Kredel.
\newblock On a {Java} {Computer} {Algebra} {System}, its performance and
  applications.
\newblock In {\em Special Issue of PPPJ 2006}, page (in print). Science of
  Computer Programming, Elsevier, 2008.

\bibitem{Odersky:2003}
{Martin Odersky}.
\newblock The {Scala} programming language.
\newblock Technical report, http://www.scala-lang.org/, accessed 2008, April,
  2003-2008.

\bibitem{Niculescu:2003}
V.~Niculescu.
\newblock A design proposal for an object oriented algebraic library.
\newblock Technical report, Studia Universitatis "Babes-Bolyai", 2003.

\bibitem{Niculescu:2004}
V.~Niculescu.
\newblock {OOLACA}: an object oriented library for abstract and computational
  algebra.
\newblock In {\em OOPSLA Companion}, pages 160--161. ACM, 2004.

\bibitem{Parisse:2008}
B.~Parisse.
\newblock {Giac/Xcas}, a free computer algebra system.
\newblock Technical report, University of Grenoble, 2008.

\bibitem{Platzer:2005}
A.~Platzer.
\newblock The {Orbital} library.
\newblock Technical report, University of Karlsruhe, http://
  www.functologic.com/, 2005.

\bibitem{Stein:2005}
W.~Stein.
\newblock {\em {SAGE} {M}athematics {S}oftware ({V}ersion 2.7)}.
\newblock The SAGE~Group, 2007.
\newblock http://www.sagemath.org, accessed 2007, November.

\bibitem{Whelan:2003}
C.~Whelan, A.~Duffy, A.~Burnett, and T.~Dowling.
\newblock A {Java} {API} for polynomial arithmetic.
\newblock In {\em Proc. PPPJ'03}, pages 139--144, New York, 2003. Computer
  Science Press.

\bibitem{Matsumo:1995}
{Yukihiro Matsumo}.
\newblock {Ruby} a dynamic, open source programming language with a focus on
  simplicity and productivity.
\newblock Technical report, http://www.ruby-lang.org/, accessed 2008, April,
  1995-2008.

\bibitem{Zippel:2001}
R.~Zippel.
\newblock {Weyl} computer algebra substrate.
\newblock In {\em Proc. DISCO '93}, pages 303--318. Springer-Verlag Lecture
  Notes in Computer Science 722, 2001.

\end{thebibliography}

\end{document}